\input epsf

% =======================LETRAS HUECAS============================
\newfam\msbfam
\font\twlmsb=msbm10 at 12pt
\font\eightmsb=msbm10 at 8pt
\font\sixmsb=msbm10 at 6pt
\textfont\msbfam=\twlmsb
\scriptfont\msbfam=\eightmsb
\scriptscriptfont\msbfam=\sixmsb

\centerline{\bf ENTROPY OF A RINDLER OBSERVER} 

\

\centerline{O. Brauer}
\centerline{\it Facultad de Ciencias, Universidad Nacional Aut\'onoma de M\'exico}
\centerline{\it Circuito Exterior, Ciudad Universitaria, 04510, M\'exico D. F., M\'exico} 

\centerline{E. Kirchuk}
\centerline{\it Departamento de Matem\'aticas, CBC, Universidad de Buenos Aires}
\centerline{\it Ciudad Universitaria, Av. Cantilo S/N, Pab. III, CP 1428, Ciudad de Buenos Aires, Argentina}

\centerline{L. Raviola}
\centerline{\it Instituto de Industria, Universidad Nacional de General Sarmiento}
\centerline{\it J. M. Guti\'errez 1153, CP 1630, Los Polvorines, Pcia. de Buenos Aires, Argentina}

\centerline{M. Socolovsky}
\centerline{\it  Instituto de Ciencias Nucleares, Universidad Nacional Aut\'onoma de M\'exico}
\centerline{\it Circuito Exterior, Ciudad Universitaria, 04510, M\'exico D. F., M\'exico} 

\

{\bf Abstract.} {\it We compute the entropy of a Rindler particle-detector (observer) in the presence of a quantum field in the Minkowski vacuum state; due to the Unruh effect, the observer is immersed in a thermal bath at a temperature proportional to its proper acceleration. }

\

{\bf 1.} Hyperbolic motion [1], [2], is the simplest non trivial motion of a classical massive particle in the context of special relativity. It consists, in the instantaneous rest frame, of a rectilinear accelerated motion with constant (proper) acceleration $\alpha$. Restricting ourselves for simplicity to the $(ct,x)$-plane of Minkowski spacetime, the trajectory of a particle which at $t=0$ passes through the point $(ct,x)=(0,c^2/\alpha)$ with zero velocity is given by $$x^2-c^2t^2=(c^2/\alpha)^2 \eqno{(1)}$$ with velocity $$\dot{x}={{\alpha t}\over{\sqrt{1+(\alpha t/c)^2}}}.\eqno{(2)}$$ It is clear that the lines $x=\pm ct$, respectively are the future and past horizons for the particle: no signal can reach to it if emitted at $(ct,x)$ with $ct>x$, and the particle can not send information to any point $(ct,x)$ with $ct<x$. (See Fig. 1.)

\

\centerline{\epsfxsize=65ex\epsfbox{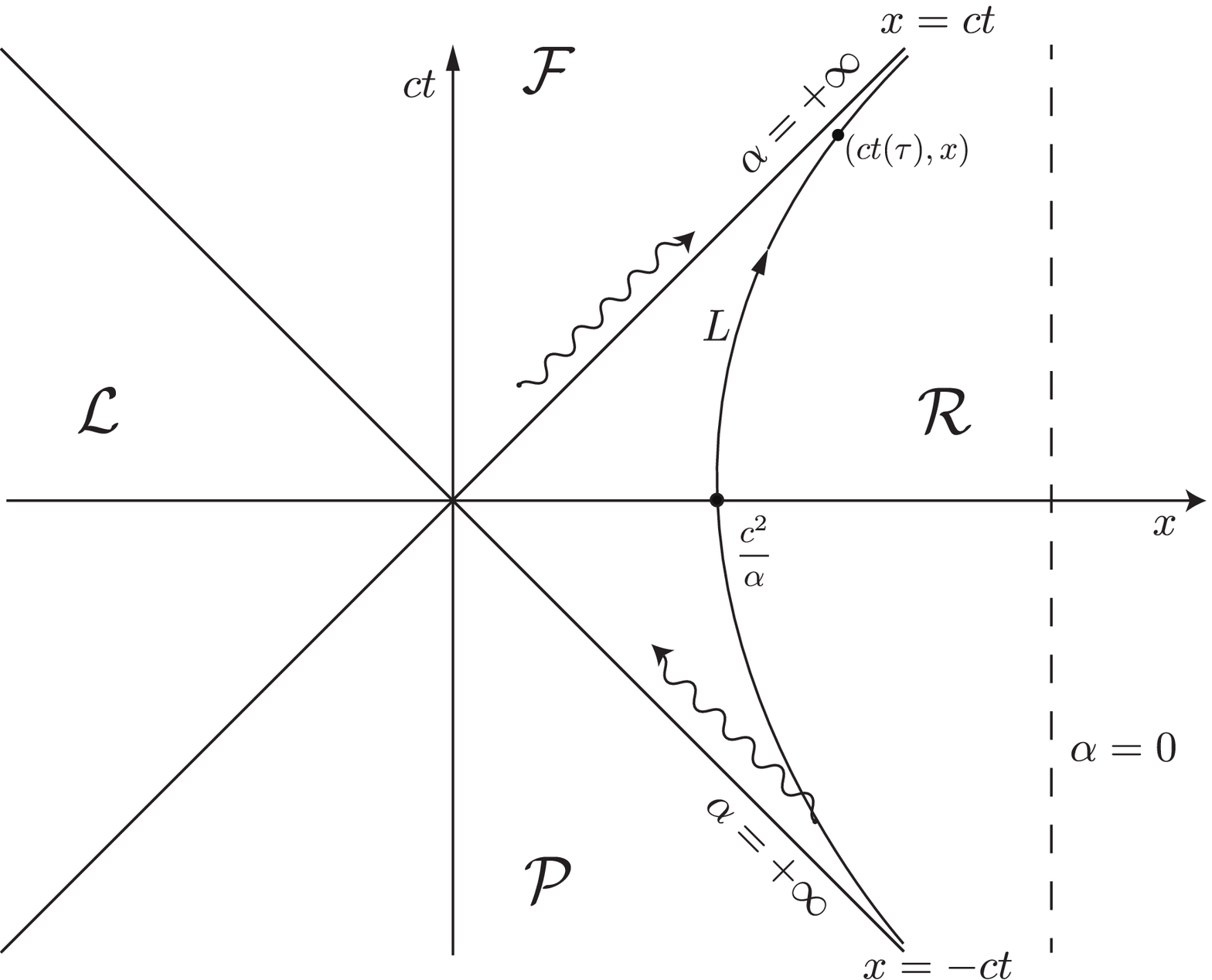}}

\

\centerline{Fig. 1}

\

The motion extends from $x=+\infty$ at $t=-\infty$ to $x=+\infty$ at $t=+\infty$. Asymptotically, the velocity tends to $-c$ and $+c$ respectively. The right region interior to the horizons is called the right Rindler wedge or, more simply, Rindler space; it is denoted by ${\cal R}$ [3]. As $\alpha \to +\infty$, the hyperboles tend to the horizons, so ``the proper acceleration of a photon can be taken to be infinite" [4]; while as $\alpha \to 0_+$, $c^2/\alpha\to +\infty$ and the particle is at rest at $x=+\infty$ (dashed line in Fig. 1). ${\cal F}$, ${\cal L}$, and ${\cal P}$ in Fig. 1 are the future, left, and past wedges. The relation between proper time $\tau$ and coordinate time $t$ is $$t={{c}\over{\alpha}}Sh({{\alpha\tau}\over{c}}),\eqno{(3)}$$ and the acceleration vector is $$\alpha^\mu=(\alpha^0,\alpha^1)={u^\mu}^\prime=\gamma\dot{u}^\mu=\gamma(c\dot{\gamma},{{d}\over{dt}}(\gamma\dot{x}))\eqno{(4)}$$ with $u^\mu=(c\gamma,\gamma\dot{x})$, $\gamma=1/\sqrt{1-{{\dot{x}^2}\over{c^2}}}$, ${u^\mu}^\prime={{du^\mu}\over{d\tau}}$, $\dot{u}^\mu={{du^\mu}\over{dt}}$. From $$-\eta_{\mu\nu}\alpha^\mu\alpha^\nu=\alpha^2\eqno{(5)}$$ with $\eta_{\mu\nu}=\pmatrix{1 & 0 \cr 0 & -1\cr}$, one obtains $$\alpha^1=\sqrt{\alpha^2+{{1}\over{c^2}}{{(\dot{x}\ddot{x})^2}\over{(1-\dot{x}^2/c^2)^4}}}=\alpha Ch({{\alpha\tau}\over{c}}).\eqno{(6)}$$

\

{\bf 2.} In 1976, Unruh [5] found that, if a quantum field in its vacuum state is present in the Minkowski spacetime, the accelerated particle (``Rindler observer" or ``particle-detector") would detect that vacuum as a thermal bath or radiation with absolute temperature $$T_U\equiv T={{\hbar\alpha}\over{2\pi k_Bc}} \ (={{\alpha}\over{2\pi}} \ in \ natural \ units)\eqno{(7)}$$ with $\hbar$ and $k_B$ the Planck and Boltzmann constants. This means that to have a temperature of $1 \ K$ one needs an acceleration of approximately $10^{18}{{m}\over{sec^2}}$! For the light cones $T\to +\infty$, and for a particle at rest or in uniform motion $T=0$. Assuming with Unruh that the accelerated body comes to equilibrium with the radiation [6], one can consider each hyperbolic motion as a {\it reversible isotherm} at temperature $T$. 

\

{\bf 3.} In 1987, Laflamme [7] computed the entropy of the Rindler wedge ${\cal R}$, and found an infinite result: one quarter of the area of the event horizon (similarly to what occurs for the Schwarzschild black hole [8]). The claim was, however, that the entropy per unit area is finite. The result fits together with the infinite value of the temperature at the horizon; however, it says nothing with respect to an entropy $S(T)$ which can be naturally associated to each accelerated motion, corresponding to a temperature $T$. This leads to the possibility of studying some aspects of the thermodynamics of the particle-detector.

\

{\bf 4.} Our starting point is the relation between the variation of internal energy, absorbed heat, and work done on a system, as given by the first and second laws of thermodynamics: $$dU=\delta Q+\delta W.\eqno{(8)}$$ For a reversible process, $\delta Q=TdS$, and $\delta W=f^1dx$, where $f^1=m\alpha^1$ ($m$ is the mass of the particle). The internal energy is, in this case, the kinetic energy $E=\sqrt{p^2c^2+m^2c^4}$, with $p={{m\dot{x}}\over{\sqrt{1-\dot{x}^2/c^2}}}$, which gives $$E=mc^2\sqrt{1+({{\alpha t}\over{c}})^2}=mc^2Ch({{\alpha\tau}\over{c}}).\eqno{(9)}$$ Then, $$T{{dS}\over{dt}}={{dE}\over{dt}}-f^1\dot{x}, \eqno{(10)}$$ and using ${{d}\over{dt}}={{1}\over{Ch({{\alpha\tau}\over{c}})}}{{d}\over{d\tau}}$, one obtains $${{dS}\over{d\tau}}={{mc\alpha}\over{T}}Sh({{\alpha\tau}\over{c}})(1-Ch({{\alpha\tau}\over{c}}))={{2\pi k_Bmc^2}\over{\hbar}}Sh({{2\pi k_B}\over{\hbar}}T\tau)(1-Ch({{2\pi k_B}\over{\hbar}}T\tau))$$ $$(=2\pi mSh(2\pi T\tau)(1-Ch(2\pi T\tau)) \ in \ natural \ units).\eqno{(11)}$$ Integrating this expression between $\tau_1$ and $\tau_2$, the variation of the entropy along the isotherm is $$\Delta S(\tau_2,\tau_1;T)=S(\tau_2,T)-S(\tau_1,T), \eqno{(12)}$$ with $$S(\tau_i,T)={{mc^2}\over{T}}(Ch({{2\pi k_B}\over{\hbar}}T\tau_i)-{{1}\over{2}}Sh^2({{2\pi k_B}\over{\hbar}}T\tau_i))$$ $$(={{m}\over{T}}(Ch(2\pi T\tau_i)-{{1}\over{2}}Sh^2(2\pi T\tau_i)) \ in \ natural \ units), \ i=1,2.\eqno{(13)}$$ Since $$S(-\tau_i,T)=S(\tau_i,T),\eqno{(14)}$$ then $$\Delta S(\tau,-\tau;T)=0;\eqno{(15)}$$ in particular $\Delta S(-\infty,+\infty;T)=0$: for any finite part of the motion from $-\tau$ to $+\tau$ (or for the infinite total motion from $\tau=-\infty$ to $\tau=+\infty$) the isothermal completes a {\it cycle}, and the entropy, being a state function, returns to its initial value. 

\

In particular, $$\Delta S(\tau,0;T)={{mc^2}\over{T}}(Ch({{2\pi k_B}\over{\hbar}}T\tau)-{{1}\over{2}}Sh^2({{2\pi k_B}\over{\hbar}}T\tau)-1).\eqno{(16)}$$

\

For any fixed $T>0$, one can verify the following limit behavior: 

\

$\tau_2$ fixed, $\tau_1\to -\infty$ (past horizon): $$\Delta S(\tau_2,\tau_1;T)\sim {{mc^2}\over{8T}}e^{2\vert\lambda\vert}\to +\infty, \ \lambda={{2\pi k_B}\over{\hbar}}T\tau_1, \eqno{(17)}$$ maximal entropy, minimum order;

\

$\tau_1$ fixed, $\tau_2\to +\infty$ (future horizon): $$\Delta S(\tau_2,\tau_1;T)\sim -{{mc^2}\over{8T}}e^{2\eta}\to -\infty, \ \eta={{2\pi k_B}\over{\hbar}}T\tau_2, \eqno{(18)}$$ minimal entropy, maximum order. 

\

If we call $$h(y)=Ch(y)-{{1}\over{2}}Sh^2(y), \ y={{2\pi k_B}\over{\hbar}}T\tau, \eqno{(19)}$$ then $$\Delta S(\eta,\lambda;T)={{mc^2}\over{T}}(h(\eta)-h(\lambda))\eqno{(20)}$$ which, for $\lambda=\lambda_0=0$ ($\tau_1=0$) and $\eta>0$ ($\tau_2>0$) has the behavior plotted in Fig. 2.

\

\centerline{\epsfxsize=70ex\epsfbox{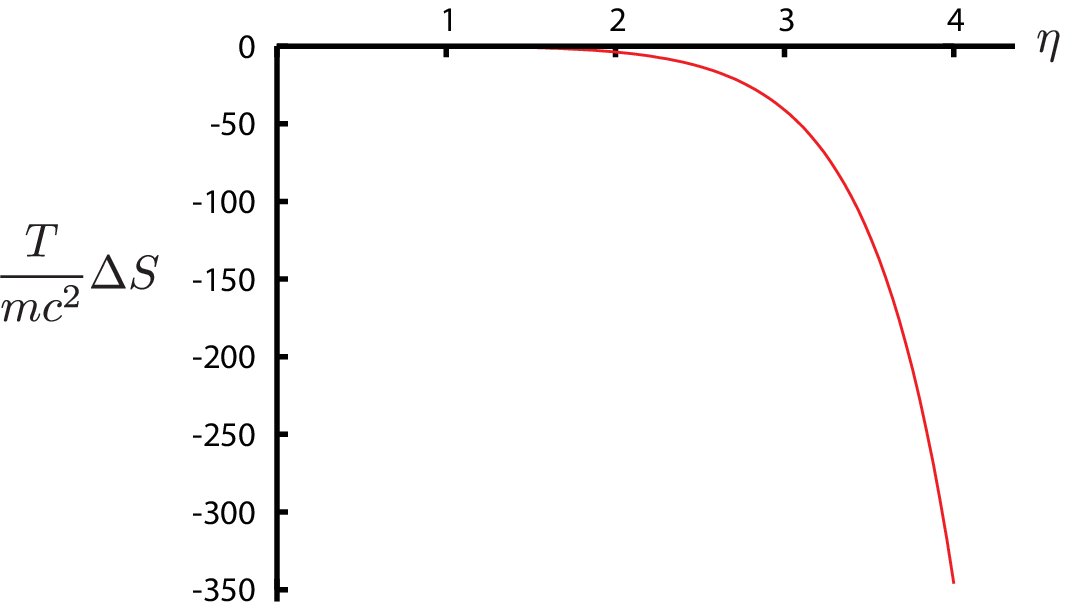}}

\

\centerline{Fig. 2}

\

For arbitrary $\tau>0$, the expansion of $Ch(\eta)-{{1}\over{2}}Sh^2(\eta)-1$ near $\eta=0$ ($T=0$) leads to\ $$\Delta S(\tau,0;T)\cong -2\pi^4{{c^2k_B^4}\over{\hbar^4}}m\tau^4T^3+O(T^4).\eqno{(21)}$$ So, $\Delta S\to 0_-$ as $T\to 0_+$, i.e. the entropy remains constant for the inertial particle. In the other extreme, for large $T$ but fixed $\tau >0$ i.e. near the future horizon, $$\Delta S(\tau,0;T)\cong -{{mc^2}\over{8T}}e^{4\pi{{k_B}\over{\hbar}}T\tau}\to -\infty \ \ as \ \ T\to +\infty.\eqno{(22)}$$

\

In terms of the coordinate time $t$, from (3), (13) is given by $$S(t,T)={{mc^2}\over{T}}(\sqrt{1+{{4\pi^2k_B^2}\over{\hbar^2}}t^2T^2}-{{2\pi^2k_B^2}\over{\hbar^2}}t^2T^2),\eqno{(23)}$$ and so $$\Delta S(t,0;T)={{mc^2}\over{T}}(\sqrt{1+{{4\pi^2k_B^2}\over{\hbar^2}}t^2T^2}-{{2\pi^2k_B^2}\over{\hbar^2}}t^2T^2-1).\eqno{(24)}$$

\

{\bf 5.} One can give a geometric interpretation of the result (16) analogous to that given by Laflamme to the Rindler wedge, but different in the sense that now it is not the horizon area the relevant quantity. In the present case, the only spatial geometric characteristic of the hyperbolic motion is the spatial length $L$  traveled  by the observer during some interval of its proper time, say between 0 and $\tau$. An easy calculation leads to $$L={{c^2}\over{\alpha}}(Ch({{\alpha\tau}\over{c}})-1)\eqno{(25)}$$ and therefore, in terms of $L$ and $T$, (16) becomes $$\Delta S(L,0;T)=-2\pi^2({{k_B}\over{\hbar}})^2mL^2T\eqno{(26)}$$ which, per unit mass, in terms of the acceleration, and in natural units gives $${{\Delta S(L,0;T)}\over{m}}=-\alpha A,\eqno{(27)}$$ where $A=\pi L^2$. Though $A$ is the area of a disk, we can not at the moment give to it a physical interpretation. As a comparison with the Schwarzschild black hole, in this case the radious of the disk (equatorial cross section) is the Schwarzschild radious.

\

{\bf Acknowledgments.} M.S. thanks for hospitality to the Instituto de Ciencias de la Universidad Nacional de General Sarmiento, Pcia. de Buenos Aires, Argentina, where part of this work was done. This work was partially supported by the project PAPIIT IN101711-2, DGAPA, UNAM, M\'exico.

\

{\bf References.}

\

[1] Landau, L. D. and Lifshitz, E. M. (1975). {\it The Classical Theory of Fields, Course of Theoretical Physics, Vol. 2}, Elsevier, Amsterdam; p. 24.

\

[2] Socolovsky, M. (2013). Rindler Space and Unruh Effect, arXiv: gr-qc/1304.2833.

\

[3] Rindler, W. (1966). Kruskal Space and the Uniformly Accelerated Frame, {\it American Journal of Physics} {\bf 34}, 1174-1178.

\

[4] Rindler, W. (2006). {\it Relativity: Special, General, and Cosmological}, Oxford University Press; p. 71.

\

[5] Unruh, W. G. (1976). Notes on black-hole evaporation, {\it Physical Review D} {\bf 14}, 870-892.

\

[6] Unruh, W. G. (1992). Thermal bath and decoherence of Rindler spacetimes, {\it Physical Review D} {\bf 46}, 3271-3277.

\

[7] Laflamme, R. (1987). Entropy of a Rindler wedge, {\it Physics Letters B} {\bf 196}, 449-450.

\

[8] Hawking, S. W. (1975). Particle Creation by Black Holes, {\it Communications in Mathematical Physics} {\bf 43}, 199-220; erratum: ibid. {\bf 46}, 206.

\

\

\

\

\

\

\

\

\

\

\

\

\

\

\

\

\

\

\

\

e-mails: brauer@ciencias.unam.mx; kirchuk@df.uba.ar; lraviola@ungs.edu.ar; 

socolovs@nucleares.unam.mx

\end